\documentclass{article}
\usepackage{spconf,amsmath,graphicx}
\usepackage{pgf}
\usepackage{subfigure}
\usepackage{rotating} 
\usepackage{bm}
\usepackage{epsf,subfigure}
\usepackage{graphicx}
\usepackage{algorithm,algorithmic}
\usepackage{cite}
\usepackage{latexsym,amssymb,amsmath}
\usepackage{verbatim}
\usepackage{media9}
\usepackage{array}
\usepackage{tabularx,booktabs}
\usepackage[font=small,skip=2pt]{caption}

\newcommand\numberthis{\addtocounter{equation}{1}\tag{\theequation}}
\newcommand{\norm}[1]{\left\|#1\right\|}

\def\bD{\ensuremath{{\bf D}}}
\def\bd{\ensuremath{{\bf d}}}
\def\bX{\ensuremath{{\bf X}}}
\def\bY{\ensuremath{{\bf Y}}}

\def\bA{\ensuremath{{\bf A}}}

\def\by{\ensuremath{{\bf y}}}
\def\bI{\ensuremath{{\bf I}}}

\title{Analysis-synthesis Model Learning with Shared Features: A New Framework for Histopathological Image Classification}
\name{Xuelu Li$^1$, Vishal Monga$^1$, U. K. Arvind Rao$^2$\thanks{Research is supported by Research Scholar Grant from the American Cancer Society (RSG-16-005-01) to AR.}}
\address{$^1$Department of Electrical Engineering, Pennsylvania State University, USA \\
	$^2$Department of Bioinformatics and Computational Biology, Radiation Oncology,\\ MD Anderson Cancer Center, Houston, TX, USA}
\usepackage{etoolbox}
\patchcmd{\thebibliography}
{\settowidth}
{\setlength{\parsep}{0pt}\setlength{\itemsep}{0pt plus 0pt}\settowidth}
{}{}
\apptocmd{\thebibliography}
{\small}
{}{}

\begin{document}
\pretolerance=10000
\tolerance=2000 
\emergencystretch=10pt
\maketitle
\vspace{-0.3cm}
\begin{abstract}
\vspace{-0cm}
\noindent Automated histopathological image analysis offers exciting opportunities for the early diagnosis of several medical conditions including cancer. There are however stiff practical challenges: 1.) discriminative features from such images for separating diseased vs. healthy classes are not readily apparent, and 2.) distinct classes, e.g.\ healthy vs. stages of disease continue to share several geometric features. We propose a novel Analysis-synthesis model Learning with Shared Features algorithm (ALSF) for classifying such images more effectively. In ALSF, a joint analysis and synthesis learning model is introduced to learn  the classifier and the feature extractor at the same time. In this way, the computation load in patch-level based image classification can be much reduced. Crucially, we integrate into this framework the learning of a low rank shared dictionary and a shared analysis operator, which more accurately represents both similarities and differences in histopathological images from distinct classes.  ALSF is evaluated on two challenging databases: (1)  kidney tissue images provided by the Animal Diagnosis Lab (ADL) at the Pennsylvania State University and (2) brain tumor images from The Cancer Genome Atlas (TCGA) database. Experimental results confirm that ALSF can offer benefits over state of the art alternatives.
\end{abstract}

\vspace{-0.3cm}
\section{Introduction}
\vspace{-0.2cm}
\label{sec:intro}
The automated classification of histopathological images has gained significant attention recently because of the promise of achieving early diagnosis for many diseases including cancer\cite{MHisto2009,USHIRC2013,USimul2014,NClass2013,VHisto2016,TDFDL2015}. These advanced image analysis
methods have been developed with three main purposes of
(i) relieving the workload on pathologists by sieving out obviously diseased and also healthy cases, which allows specialists to spend more time on more sophisticated cases; (ii) reducing inter-expert variability; and (iii) understanding the underlying reasons for a specific diagnosis to aid/enhance clinical understanding.

From a practical algorithm design viewpoint, the presence of rich geometric structures contained in the histopathological images provide researchers with both opportunities and challenges. In particular, the diversity of meaningful information and well as shared common properties of different tissue images make it hard to extract the essential discriminative features to facilitate the classification process. The computation load can also be quite demanding when dealing with large spatial size and high resolution images. As a result, most  existing algorithms operate and examine image structure at the patch level \cite{USHIRC2013,NClass2013,VHisto2016}.

While it is known that the literature on histopathological image classification is broad - see \cite{MHisto2009,VHisto2016} for a review, recent work has demonstrated the success of sparsity based classification and dictionary learning methods \cite{USimul2014,NClass2013,TDFDL2015}.
The traditional sparsity based classification algorithms take the advantage of the synthesis model of the signal expression. That is to say, they express the observed data as a linear combination of the columns from the dictionary $\bD$, and try to find the class which yields the least reconstruction error under a sparsity constraint on recovery coefficients. In more recent work, these ideas have been extended to what is known as an analysis-synthesis model \cite{HAnaly2013}. That is, a linear analysis operator $\bA$ which determines the sparse code from the images and a more typical synthesis dictionary $\bD$, which is multiplied with the sparse code to yield images, are learned simultaneously. When performed in a classification context by using class specific analysis and synthesis dictionaries, the authors in \cite{SProj2014} demonstrate that the analysis-synthesis model can outperform traditional sparsity based classification methods as in \cite{JohnRobust2009} with a lower computational cost. Further, for histopathological images, the analysis model is pretty suitable since most sparsity-based classification algorithms are applied on the patch level. In order to predict the label of a test image, thousands of patches from the image should be classified. In most sparsity based classification methods viz. \cite{USimul2014,NClass2013,TDFDL2015}, a sparsity constrained optimization should be solved to determine class labels. In ALSF, this step is computationally lot simpler, involves no significant optimization but instead some simple linear algebraic operations.  

Another outstanding open challenge in histopathological image classification is that local image regions even from distinct classes share common features. We contend hence that a shared analysis and synthesis dictionary in addition to discriminative class specific dictionaries should be included in the model. There has been recent advocacy for the use of shared dictionaries in traditional sparse coding set-ups \cite{ZLearn2012,TFast2017}. In this work, we develop a new optimization problem which learns shared as well as discriminative analysis and synthesis dictionaries under appropriate problem specific constraints. We also develop efficient optimization methods to solve the aforementioned optimization problem.
\vspace{-0.3cm}
\section{Analysis-synthesis model based shared feature learning}
\vspace{-0.3cm}
\label{sec:format}
\subsection{Joint Analysis-synthesis learning based classification}
\setlength{\belowdisplayskip}{1pt} \setlength{\belowdisplayshortskip}{1pt}
\setlength{\abovedisplayskip}{1pt} \setlength{\abovedisplayshortskip}{1pt}
\label{ssec:subhead}
Assume that we have images from $C$ different classes to form the observed data $\bY=[\bY_1,\dots,\bY_c,\dots\bY_C]$,  where$\bY_c=[\by^{1}_c,\by^{2}_c,\dots,\by^{N_c}_c]\in\mathbb{R}^{d\times N_c}$ is the data corresponding to the $c^{th}$ class, and $\by^{i}_c,i=[1,\dots,N_c]$, $c=[1,\dots,C]$ is the vectorization of $i$th image sample (patch) from class $c$.  
In the traditional discriminative dictionary learning model, a synthesis dictionary $\bD=[\bD_1,\dots,\bD_c,\dots\bD_C]$ is learned to obtain a linear representation of the observed data via the calculated sparse coefficient $\bX=[\bX_1,\dots, \bX_c,\dots,\bX_C]$, where  $\bD_c\in\mathbb{R}^{d\times k_c}$ is the sub dictionary corresponding to $c^{th}$ class, and $\bX_c=[\bX_{1c};\dots;\bX_{{c}^{\prime}c};\dots;\bX_{Cc}]$, where $\bX_{{c}^{\prime}c}\in\mathbb{R}^{k_{c^{\prime}}\times N_c}, c^{\prime}=[1,\dots,C]$. 

 Since for the classification problems, the discriminative information between different images ia more important than the accurate recovery of the image, an analysis operator $\bA=[\bA_1;\dots;\bA_{c};\dots;\bA_C]$, where $\bA_{c}\in\mathbb{R}^{k_{c}\times d}$ is learned to replace the sparse coding process to capture the sparse  and discriminative information of the observed data. Therefore, a basic joint analysis-synthesis learning model can be expressed as below:
\begin{align*}
	\{\bD^\ast,\bA^\ast\}=&\underset{\bD,\bA}{\arg \min}\sum^C_{c=1}\norm{\bY_c-\bD_{c}\bA_c \bY_c}^2_F+\lambda\norm{\bA_c \bY_{\bar{c}}}^2_F\\
	&s.t. \norm{\bd_k}^2_F\leq1, k=[1,\dots,k_c]\numberthis
\end{align*}
where $\bY_{\bar{c}}$ denotes the complementary data matrix of $\bY_{c}$ in the whole training data, and the term $\norm{\bA_c \bY_{\bar{c}}}^2_F$ is used to suppress the extracted value which does not belong to $c^{th}$ class. The classification decision of any image sample $\by$ will be made according to
\begin{equation}
c^{\ast} = \underset{c\in \{1\dots C\}}{\arg \min}\norm{\by-\bD_c \delta_{c}(\bA \by)}^2_F
\end{equation}
where $\delta_{c}(\bA \by)$ denotes the part of $\bA \by$ corresponding to the $c^{th}$ class. It can be observed that $\bD$ and $\bA$  can be regarded as a classifier and a feature extractor, respectively. hence, the sparse coding process (which typically involves solving a relatively expensive optimization problem in the test/classification step) is replaced by a simple multiplication between $\bA$ and $\by$. See \cite{SProj2014} for more details.

\subsection{Learning with shared features}
\begin{figure}[t]
	\centering
	\begin{minipage}[b]{0.80\linewidth}
		\centering
		\centerline{\includegraphics[width=7.8cm,height=3.3cm]{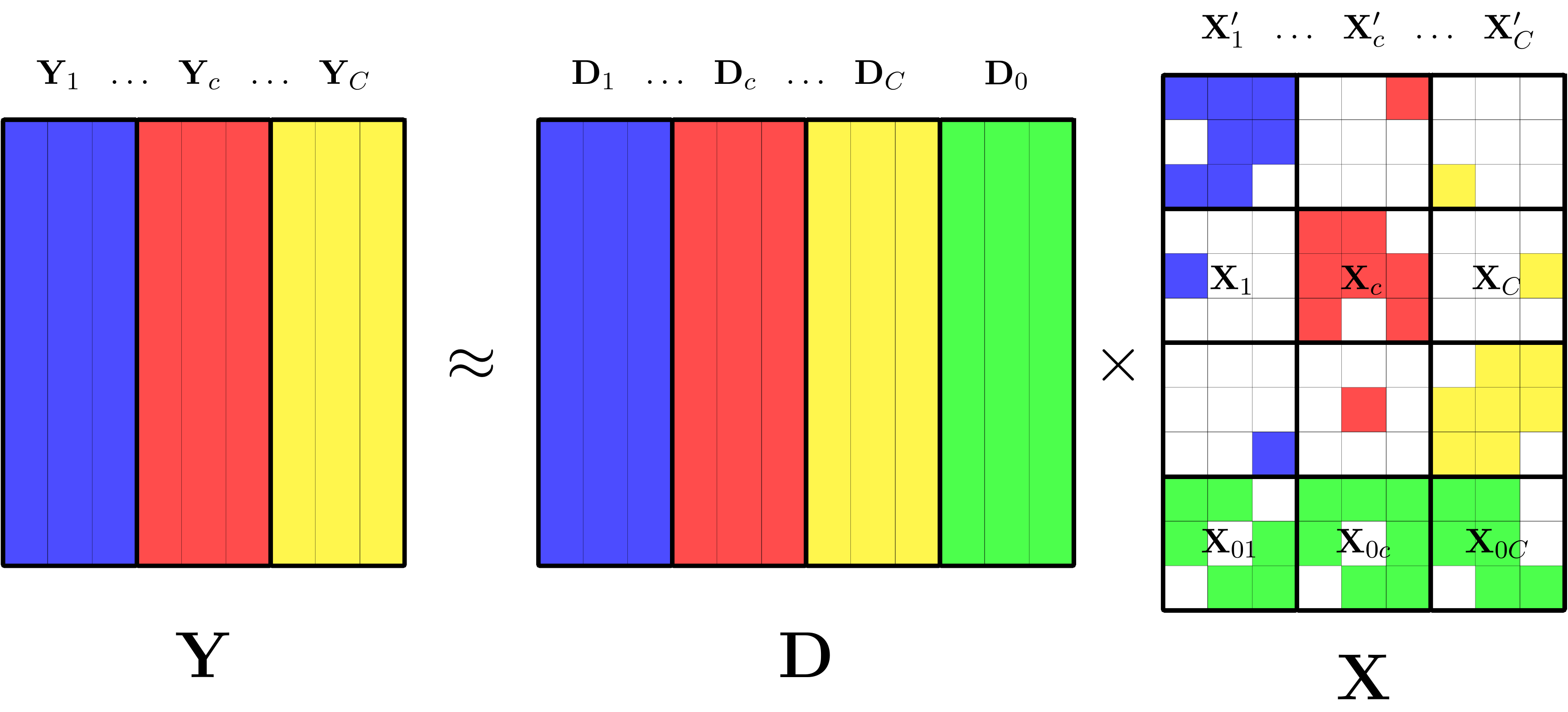}}
        \caption{Synthesis model with shared features}
        \label{Fig:synthesis}
	\end{minipage}
		\begin{minipage}[b]{0.80\linewidth}
			\centering
			\centerline{\includegraphics[width=7.8cm,height=3.3cm]{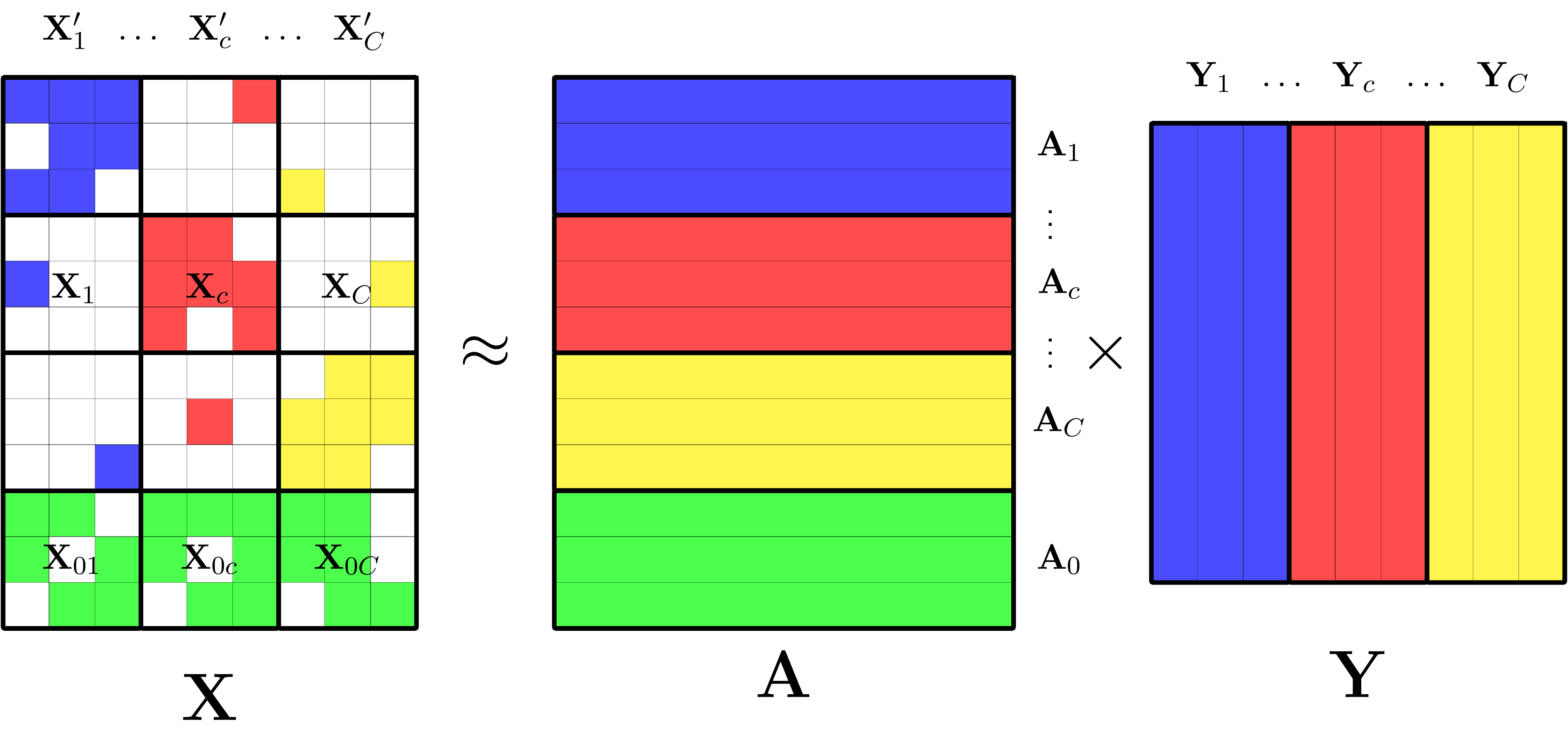}}
			\caption{Analysis model with shared features}
			\label{Fig:analysis}
		\end{minipage}
		\vspace{-0.5cm}
\end{figure}

\setlength{\belowdisplayskip}{1pt} \setlength{\belowdisplayshortskip}{1pt}
\setlength{\abovedisplayskip}{1pt} \setlength{\abovedisplayshortskip}{1pt}
\label{sec:pagestyle}
Despite the fact that there exists discriminative information, the histopathology image samples from different classes may still strongly be correlated with each other. Based on this fact, in our ALSF algorithm, we not only learn a class specific dictionary $\bD_c$ and corresponding analysis operator $\bA_c$ but also a shared dictionary $\bD_0\in\mathbb{R}^{d\times k_{0}}$ and shared analysis operator $\bA_0\in\mathbb{R}^{k_{0}\times d}$ at the same time. Our premise is that by appropriately capturing the shared or common part, the discriminative information from different classes extracted by $\bA_c$ can now be better represented by $\bD_c$. The relation between $\bD_c$ and $\bD_0$ and the relation between $\bA_c$ and $\bA_0$ is illustrated in  Fig.\ref{Fig:synthesis} and Fig.\ref{Fig:analysis}, respectively, where $\bX^\prime_c=[\bX_c;\bX_{0c}]$.

We now formulate an optimization problem to learn the required parameters for the proposed algorithm. 

For the classifier part:  
\begin{align*}
{\bD^{\ast}} = \underset{\bD}{\arg \min}&\sum_{c=1}^{C}\norm{\bY_c-\bD_{c}\bX_{cc}-\bD_{0}\bX_{0c}}^2_F+\eta\norm{\bD_0}_\ast\\
& s.t \norm{\bd_k}^2_F\leq1, k=1,\dots,k_c\numberthis
\end{align*}
where $\norm{\bD_0}_\ast$ is the popular nuclear norm surrogate that minimizes the rank of $\bD_0$. This is crucial because unless the rank of $\bD_0$, its synthesis ability (spanning subspace) could extend to capture even discriminative features which is undesirable.

{ For the feature extraction/analysis part:}
\begin{align*}
{\bA^{\ast}} 
= \underset{\bA}{\arg \min}&
\sum_{c=1}^{C}\frac{1}{N_{\bar{c}}}\norm{\bA_c\bY_{\bar{c}}}^2_F+\lambda_1\norm{\bA_0(\bY_c-\bY^{m})}^2_F\\
& s.t \norm{\bA}^2_F\leq\sigma, \sigma>0\numberthis
\end{align*}
where $\norm{\bA}^2_F\leq\sigma$ is used to suppress the coefficient values of $\bA$, $N_{\bar{c}}$ denotes the number of image samples which do not belong to the $c^{th}$ class, $\bY^{m}\in\mathbb{R}^{d\times N_c}$ is the matrix whose each column is the average value of the observed data within the training data. The second term $\norm{\bA_0(\bY_c-\bY^{m})}^2_F$ is trying to make the extracted features from different classes corresponding to the common properties to be similar with each other. Both the classifier and the feature extractor should satisfy the condition that $\bX_{ij}=\bA_i\bY_j, i=[0,\dots,C], j=[1,\dots,C]$.
Let
\begin{align*}
f(\bD_c,\bA_c,\bD_0,\bX_{cc},\bX_{0c})=&\norm{\bY_c-\bD_{c}\bX_{cc}-\bD_{0}\bX_{0c}}^2_F\\&+\eta\norm{\bD_0}_\ast\numberthis
\end{align*}
\begin{align*}
g(\bA_c,\bA_0) 
=\frac{1}{N_{\bar{c}}}\norm{\bA_c\bY_{\bar{c}}}^2_F+\lambda_1\norm{\bA_0(\bY_c-\bY^{m})}^2_F\numberthis
\end{align*}
Then we can obtain the complete form of the optimization problem as below:
\begin{align*}
\label{eq:mainOpt}
&\{\bD^{\ast}_c,\bA^{\ast}_c,\bD_0^{\ast},\bA_0^{\ast},\bX_{cc}^{\ast},\bX_{0c}^{\ast}\}
=\\&\underset{\bD_c,\bA_c,\bD_0,\bA_0,\bX_{cc},\bX_{0c}}{\arg \min}
\sum_{c=1}^{C}\{f(\bD_c,\bA_c,\bD_0,\bX_{cc},\bX_{0c})\\ &~~~~~~~+\tau g(\bA_c,\bA_0)+\tau\lambda_2\norm{\bX_{cc}-\bA_c\bY_c}^2_F
\\&~~~~~~~~~~~~~~~~~~~~~~~~ +\tau\lambda_3\norm{\bX_{0c}-\bA_0\bY_c}^2_F\}\\
&~~~~~~~~~~~~~~~~~~~ s.t \norm{\bd_k}^2_F\leq1, \norm{\bA}^2_F\leq\sigma\numberthis
\end{align*}
\vspace{-.19in}
\subsection{Efficient Optimization Algorithms}
\setlength{\belowdisplayskip}{1pt} \setlength{\belowdisplayshortskip}{1pt}
\setlength{\abovedisplayskip}{1pt} \setlength{\abovedisplayshortskip}{1pt}
\label{sec:pagestyle}

The problem in (\ref{eq:mainOpt}) is challenging but can be solved efficiently by breaking into several sub-problems.

Initialization: $\bD,\bD_0$\cite{ZLearn2012}, $[\bA;\bA_0]=[\bD, \bD_0]^{\dagger}$,
where $(\cdot)^{\dagger}$ represents the pseudoinverse:
%

\begin{align*}
\{&\bX_{cc}^\ast,\bX_{0c}^\ast\}=\underset{\bX_{cc}}{\arg \min}\norm{\bY_c-\bD_{c}\bX_{cc}-\bD_{0}\bX_{0c}}^2_F\\&+\tau\lambda_2\norm{\bX_{cc}-\bA_c\bY_c}^2_F
+\tau\lambda_3\norm{\bX_{0c}-\bA_0\bY_c}^2_F \numberthis
\end{align*}
\begin{align*}
\bA_c^\ast=\underset{\bA_{c}}{\arg \min}\frac{1}{N_{\bar{c}}}\norm{\bA_c\bY_{\bar{c}}}^2_F+&\lambda_2\norm{\bX_{cc}-\bA_c\bY_c}^2_F\\&+\eta_1\norm{\bA_c}^2_F\numberthis
\end{align*}
	\begin{equation}
	\bA_0^\ast=\underset{\bA_{0}}{\arg \min}\sum_{c=1}^C\norm{\bA_0(\bY_c-\bY^{m})}^2_F+\frac{\lambda_3}{\lambda_1}\norm{\bX_{0c}-\bA_0\bY_c}^2_F
	\end{equation}
	\begin{equation}
	\bD_c^{\ast}={\arg \min}\norm{\bY_c-\bD_{c}\bX_{cc}-\bD_{0}\bX_{0c}}^2_F
	\end{equation}
	\begin{equation}
	\bD_0^{\ast}=\underset{\bD_{0}}{\arg \min}\sum_{c=1}^{C}\norm{\bY_c-\bD_{c}\bX_{cc}-\bD_{0}\bX_{0c}}^2_F+\eta\norm{\bD_0}_\ast
	\end{equation}
	where $\lambda$, $\eta$, $\lambda_1$, $\lambda_2$, $\lambda_3$, $\tau$ and $\eta_1$ appeared above are regularization parameters used to balance the cost function. The values of these parameters are set in practice by a cross-validation procedure \cite{KohaviStudy1995}.
The closed form solutions of each of these sub-problems are given by:
	\begin{equation}
	\bX_{cc}^\ast=\underset{\bX_{c}}{\arg \min}\norm{
		\begin{bmatrix}
		\bY_c-\bD_0\bX_{0c}\\
		\sqrt{\tau\lambda_2}\bA_c\bY_c
		\end{bmatrix}
		-
		\begin{bmatrix}
		\bD_c\\
		\sqrt{\tau\lambda_2}\bI
		\end{bmatrix}\bX_{cc}}^2_F
	\end{equation}
	\begin{equation}
	\bX_{0c}^\ast=\underset{\bX_{0c}}{\arg \min}\norm{
		\begin{bmatrix}
		\bY_c-\bD_c\bX_{c}\\
		\sqrt{\tau\lambda_3}\bA_0\bY_c
		\end{bmatrix}
		-
		\begin{bmatrix}
		\bD_0\\
		\sqrt{\tau\lambda_3}\bI
		\end{bmatrix}\bX_{0c}
	}^2_F
	\end{equation}
	\begin{align*}
	&\bA_c^\ast=\underset{\bA_{c}}{\arg \min}\\&\norm{\bA_{c}
		\begin{bmatrix}
		\frac{1}{N_{\bar{c}}}\bY_{\bar{c}} & 
		\sqrt{\lambda_2}\bY_c & \sqrt{\eta_1}\bI
		\end{bmatrix}
		-
		\begin{bmatrix}
		\bf{0} & 
		\sqrt{\lambda_2}\bX_{cc} & \bf{0}
		\end{bmatrix}
	}^2_F\numberthis
	\end{align*}
	\begin{equation}
	\bD_{c}^\ast=\underset{\bD_{c}}{\arg \min}\norm{
		\begin{bmatrix}
		\bY_c-\bD_0\bX_{0c}
		\end{bmatrix}
		-
		\bD_c\bX_{cc}
	}^2_F
	\end{equation}
\begin{equation}
\bD_0^{\ast}=\underset{\bD_{0}}{\arg \min}\norm{
	\begin{bmatrix}
	\bY_c-\bD_{c}\bX_{cc}
	\end{bmatrix}
	\bX_{0c}^{\dagger}-\bD_{0}}^2_F+\eta\norm{\bD_0}_\ast
\end{equation}
(Solved by the singular value thresholding method \cite{CaiSingular2010}.)
\begin{align*}
\small
\bA_0^\ast=\underset{\bA_{0}}{\arg \min}\norm{\bA_{0}
	\begin{bmatrix}
	\bY_1^T \\
	\sqrt{\frac{\lambda_1}{\lambda_3}}(\bY_1-\bY^m)^T \\ \dots \\\bY_C^T \\ 
	\sqrt{\frac{\lambda_1}{\lambda_3}}(\bY_C-\bY^m)^T
	\end{bmatrix}^T
	-
	\begin{bmatrix}
	\bX_{01}^T \\ 
	{\bf 0}^T\\ \dots \\ \bX_{0C}^T\\ 
	{\bf0}^T
	\end{bmatrix}^T
}^2_F \numberthis
\end{align*}
\begin{figure}[t]
	\centering
	\begin{minipage}[b]{0.40\linewidth}
		\centering
		\centerline{\includegraphics[width=3.2cm,height=2.1cm]{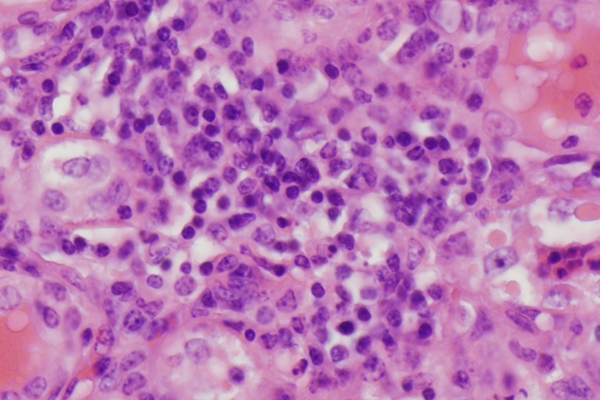}}
		\centerline{(a) Inflammed Kidney}\medskip
	\end{minipage}
	\hspace{0.5cm}
	\begin{minipage}[b]{.40\linewidth}
		\centering
		\centerline{\includegraphics[width=3.2cm,height=2.1cm]{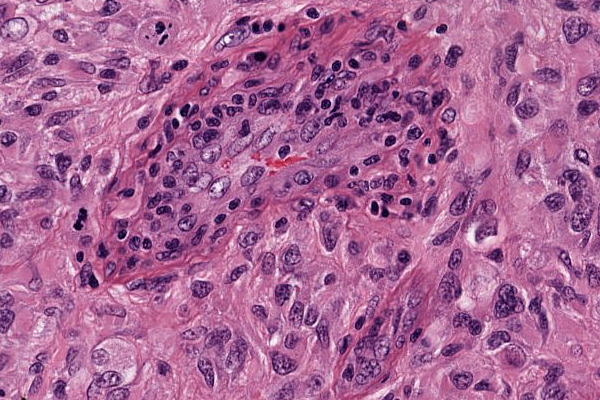}}
		\centerline{(b) MVP}\medskip
	\end{minipage}

	\begin{minipage}[b]{0.40\linewidth}
		\centering
		\centerline{\includegraphics[width=3.2cm,height=2.1cm]{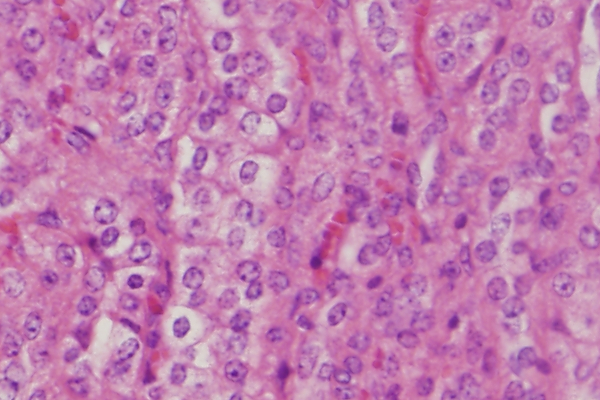}}
		\centerline{(c) Healthy Kidney}\medskip
	\end{minipage}
	\hspace{0.5cm}
	\begin{minipage}[b]{0.40\linewidth}
		\centering
		\centerline{\includegraphics[width=3.2cm,height=2.1cm]{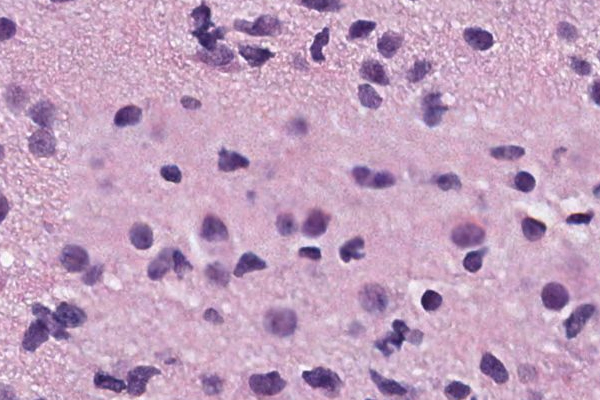}}
		\centerline{(d) NotMVP}\medskip
	\end{minipage}\\
		\vspace{-0.2cm}
	\caption{Sample images from two databases. Column 1: ADL-Kidney; Column 2: TCGA database}
	\label{fig:res}
	\vspace{-0.7cm}
	\label{Fig:image sample}
\end{figure}
where $(\cdot)^T$ represents the matrix transpose.

\vspace{-0.35cm}
\section{Experimental Results}
\vspace{-0.2cm}
\label{sec:typestyle}

The proposed ALSF algorithm is evaluated on two challenging databases (image samples are shown in Fig.\ref{Fig:image sample}). We compare against the following state-of-the-art methods: The well-known WND-CHRM features\cite{LWndch2008} combined with an SVM classifier, a host of dictionary learning methods -- Fisher Discriminative Dictionary Learning (FDDL) \cite{YFisher2011}, Discriminative Feature oriented Dictionary Learning (DFDL) \cite{VHisto2016}, and Label Consistent-KSVD (LCKSVD) \cite{ZLabel2013} which have been shown to be quite successful in related research. In each experiment, 800 20-by-20 patches are randomly extracted from training images of each class. Each learned dictionary has 400 bases per class. The down sampled test images are cropped into non-overlapping 20-by-20 (pixel) patches, and the labels of all the patches from each image are predicted through the learned feature extractor and the classifier first. The final classification decision of each image will be made according to whether the ratio of the inflammed patches in the image is above a pre-determined learned threshold (ADL) or whether there exists a connected MVP area in the image which is above a learned threshold (TCGA). The aforementioned pre-determined learned thresholds are determined by a process described in \cite{VHisto2016}.     

{\bf ADL-kidney data base} contains bovine histopathology images of kidney acquired from the Animal Diagnostics Lab, Pennsylvania State University. The dataset available for public consists of images with a size of $1360\times1024$ pixels from two classes: Inflammatory and Healthy. The images are down-sampled into $272\times205$ pixels for the purpose of computation efficiency. From the images, 40 images of each class are used as training data, and the others (138 for the inflammed and 115 for the healthy) are used as test data. The inflammed patches are mainly extracted from the center region of the training images. 
\begin{table}
	
	\setlength{\belowcaptionskip}{-0.4cm}
	\begin{center}
		\caption {Confusion matrix for the ADL-Kidney database}\label{tab:ADL-kidney}
	{\footnotesize	\begin{tabular}{ |c|c|c|c|} \hline
				class & Inflammed & Healthy & Method\\
				\hline
				           &{\bf 0.870} & 0.130 & ALSF\\
				           &0.848 & 0.162 & DFDL\\
				Inflammed  &0.797 & 0.203 & FDDL\\
				           &0.754 & 0.246 & LCKSVD\\
				           &0.703 & 0.297 & WND-CHRM\\  
				\hline
				
				           &0.145 & {\bf 0.855} & ALSF\\
				           &0.162 & 0.838 & DFDL\\
				Healthy    &0.188 & 0.812 & FDDL\\
				           &0.120 & {\bf 0.880} & LCKSVD\\
				           &0.188 & 0.812 & WND-CHRM\\  
				\hline
			
		\end{tabular} }
	\end{center}

\vspace{-0.6cm}
\end{table}

\begin{table}[t]
	\begin{center}
		\setlength{\belowcaptionskip}{-0cm}
		\caption {Confusion matrix for the TCGA database}\label{tab:TCGA}
{\footnotesize		\begin{tabular}{| c | c| c|c| } 
		\hline
			class & MVP & NotMVP & Method\\
			\hline
		             &{\bf 0.971} & 0.029 & ALSF\\
			         &{\bf 1.000} & 0.000 & DFDL\\
			MVP      &0.853 & 0.147 & FDDL\\
			         &0.794 & 0.206 & LCKSVD\\
			         &0.735 & 0.265 & WND-CHRM\\  
			\hline
	
	          	&0.049 & {\bf 0.951} & ALSF\\
		        &0.098 & 0.902 & DFDL\\
		NotMVP  &0.108 & 0.892 & FDDL\\
		        &0.078 & 0.922 & LCKSVD2\\
		        &0.226 & 0.774 & WND-CHRM\\  
		\hline
			
		\end{tabular}
	}
	\end{center}
	\vspace{-1cm}
\end{table}
\begin{table}[t]
	\setlength{\belowcaptionskip}{-0.4cm}

	\begin{center}
			\caption {Run time performances comparison}\label{tab:time}
	{\footnotesize	\begin{tabular}{ |c| c|c| } 
			\hline
			time(s) & ADL-Kidney & TCGA \\
			\hline
			ASFL &{\bf 0.058} & {\bf 0.071}\\ 
			DFDL &0.359 & 3.047\\
			FDDL &0.299 & 3.124\\ 
			LCKSVD2 &0.134 & 1.735\\ 
	
			\hline
			
		\end{tabular}
	}
	\end{center}
	\vspace{-0.8cm}
\end{table}

{\bf TCGA data base} contains brain cancer images obtained from the TCGA database provided by the National Institute of Health. The images are labeled as two classes: High grade glioma, which is indicated by the presence of the Micro Vascular Proliferation (MVP), and Low grade glioma, where no MVP region exists (denoted as NotMVP). We randomly pick 20 images from each class as the training data, and the others (34 for the MVP and 102 for the NotMVP) are used as test data. It is noted that when extracting the training patches from MVP images, we only extract the patches which are totally contained in the MVP region.

The final average classification accuracy for the two databases as shown in Table~\ref{tab:ADL-kidney} and Table~\ref{tab:TCGA} confirm the merits of the proposed ALSF. Recall, a key computational benefit of the ALSF is that no explicit sparse coding problem needs to be solved in the test/classification phase. Table~\ref{tab:time} hence compares run time of algorithms based on dictionary learning during the test or classification phase - it is readily apparent that ALSF leads to a very significant improvement.

{

\bibliographystyle{IEEEbib}
\vspace{-0.5cm}
\bibliography{strings,refs}
}

\end{document}